\documentclass[12pt]{spieman}  
\usepackage{tocloft}
\usepackage{amsmath,amsfonts,amssymb}
\usepackage{mathabx}
\usepackage{graphicx}
\usepackage{setspace}
\usepackage{float}
\usepackage{hhline}
\usepackage{lineno}
\usepackage[square,numbers]{natbib}

\newcommand{\rom}[1]{\uppercase\expandafter{\romannumeral #1\relax}}

\def\gappeq{\mathrel{ \rlap{\raise.5ex\hbox{$>$}}
                      {\lower.5ex\hbox{$\sim$}}  } }

\title{The Scientific Impact of a Noiseless Energy-Resolving Detector for a Future Exoplanet-Imaging Mission}

\author[a,b,c,*]{Alex R. Howe}
\author[a]{Christopher C. Stark}
\author[b]{John E. Sadleir}
\affil[a]{NASA Goddard Space Flight Center, 8800 Greenbelt Rd, Greenbelt, MD 20771, USA}
\affil[b]{Center for Research and Exploration in Space Science and Technology, NASA/GSFC, Greenbelt, MD 20771}
\affil[c]{Southeastern Universities Research Association, 1201 New York Avenue NW, Suite 430, Washington, DC 20005}

\cftpagenumbersoff{figure}
\cftpagenumbersoff{table} 
\begin{document} 
\maketitle

\begin{abstract}


{\normalsize Future space missions that aim to detect and characterize Earth-like exoplanets will require an instrument that efficiently measures spectra of these planets, placing strict requirements on detector performance. The upcoming Roman Space Telescope will demonstrate the performance of an electron-multiplying charge-coupled device (EMCCD) as part of the coronagraphic instrument (CGI). The recent LUVOIR and HabEx studies baselined pairing such a detector with an integral field spectrograph (IFS) to take spectra of multiple exoplanets and debris disks simultaneously. We investigate the scientific impact of a noiseless energy-resolving detector for the planned Habitable Worlds Observatory's (HWO) coronagraphic instrument. By assuming higher quantum efficiency, higher optical throughput, and zero noise, we effectively place upper limits on the impact of advancing detector technologies. We find that energy-resolving detectors would potentially take spectra of hundreds of additional exoplanets ``for free'' over the course of an HWO survey, greatly increasing its scientific yield.}

\end{abstract}

\keywords{image sensors, spectroscopy, telescopes, planets}

{\noindent \footnotesize\textbf{*}Alex R. Howe,  \linkable{alex.r.howe@nasa.gov} }

\begin{spacing}{2}   

\section{Introduction}

The Astro2020 Decadal Survey Final Report \cite{ASTRO2020} charged NASA to develop a $\sim$6-meter inscribed diameter space telescope to detect and spectrally characterize 25 potentially Earth-like planets. That mission concept is now known as the Habitable Worlds Observatory (HWO).

With apparent visual magnitudes of $\sim$31, Earth-like exoplanets will be among the faintest objects ever detected. With the telescope design we adopt in this paper, an Earth-twin at 10 pc, observed at quadrature, would have a flux of approximately 36 photons per minute over the aperture in a 20\% bandpass in V-band. This is roughly one third of the expected background noise level, assuming an exozodical disk similar to our solar system's zodiacal cloud (which itself is only one third as dense as the median exozodi \cite{ErtelExoZ}). Thus, long integration times will be needed merely to detect Earth-like exoplanets. Measuring spectra of these objects will therefore require the use of low-noise, high-quantum efficiency (QE) photon-counting detectors, potentially extending from the near ultra-violet (NUV) to near infra-red (NIR) \cite{habexfinalreport, luvoirfinalreport}.

While detectors meeting these criteria have yet to be flight proven, multiple options are in development. The Roman Coronagraph will soon demonstrate the use of an electron-multiplying charge-coupled device (EMCCD) in flight. The radiation-hardened flight EMCCD meets {\it Roman's} requirements for dQE, charge transfer efficiency, and dark current through the use of shielding, although dark current is expected to increase over time in a tail of hot pixels \cite{CCDrad}. Broadly speaking, these performance metrics are on par with what is required for future exo-Earth characterization \cite{habexfinalreport,luvoirfinalreport}. However, other performance parameters will require continued improvement. The HabEx and LUVOIR mission concept studies baselined EMCCDs with substantially improved parameters compared with {\it Roman}, with the assumption of a clock induced charge of $1.3\times10^{-3}$ counts pix$^{-1}$ frame$^{-1}$ (compared with {\it Roman's} $1.6\times10^{-2}$ counts pix$^{-1}$ frame$^{-1}$ BOL \cite{Roman}) and a quantum efficiency (QE) of 90\% at 950 nm (compared with {\it Roman's} 11.6\% \cite{Roman}).

Both the HabEX and LUVOIR mission concept studies adopted EMCCDs within an integral field spectrograph (IFS), which uses lenslets and dispersive optics to measure spectra at each point in the image plane, resulting in a spectral data cube \cite{IFSref}. Dispersing the light further reduces the signal's photon count rate per pixel, to the point that detector noise can significantly extend exposure times. As a result, both the HabEx and LUVOIR mission concepts adopted separate imaging and spectral characterization modes, such that the imaging observations would benefit from fewer noise pixels.

Here, we investigate the scientific impact of an energy-resolving detector (ERD), effectively an IFS on a chip, which offers multiple significant advantages over current IFS technologies. Such a detector would not require the dispersive optics of an IFS, thus increasing throughput while reducing the number of noise pixels. No separate imaging mode would be required, as ERDs obtain spectra at all times. ERDs also tag individual photons (for the fluxes expected in an HWO-style survey), so they do not produce false counts; instead they are limited by their measurement accuracy of photon energy (spectral resolution) \cite{Rauscher16}.

Several energy-resolving detector technologies could prove viable for HWO \cite{Rauscher16}, most notably Microwave Kinetic Inductance Detectors (MKIDs) and Transition Edge Sensor (TES) arrays. In this paper we focus on the value of ERDs in general; the development paths or engineering requirements for specific ERDs is left to future HWO trade studies. Thus, this paper is a study of the value of ERDs in general rather than individual detector technologies. However, briefly, MKIDs detect single photons and resolve their energies via their modification to the kinetic inductance of a superconducting material by freeing Cooper pairs (see e.g. \cite{MKIDref}). They have zero dark current and lend themselves well to multiplexing, but achieving the required spectral resolution of $R=140$ may be challenging \cite{Rauscher16}.
Transition Edge Sensor (TES) arrays detect single photons using the resistive transition of a superconductor, where very small changes in temperature yields large changes in resistance, providing a signal size proportional to the energy of the impinging photon \cite{TESref}. TES arrays have been shown to have no measurable degradation when exposed to radiation equivalent to a 25 year mission lifetime at L2 with Al shielding \cite{beaumont2023a,beaumont2023b}. However, all of these ERD technologies would also require substantial cooling, and all would require some level of improvement to their energy resolution to meet the $R=140$ baselined by the HabEx and LUVOIR studies \cite{habexfinalreport, luvoirfinalreport}, with MKID prototypes achieving $R\sim50$ \cite{deVisser21} and TES prototypes achieving $R\sim90$ \cite{Techport} near visible wavelengths. (See \cite{Rauscher16} for more details on possible ERD technologies.)

In this paper, we constrain the potential scientific impact of a noiseless ERD compared with the EMCCD+IFS baselined by the HabEx and LUVOIR concept studies. In Section \ref{sec:methods}, we present our methods for modeling exoplanetary systems with the ExoVista code, simulating a survey based on the LUVOIR-B design study \cite{luvoirfinalreport} and counting the resulting yield of detected planets and spectra for the two detector technologies. We discuss the results of this study in Section \ref{sec:results}, and we summarize our conclusions in \ref{sec:conclusion}.

\section{Methods}
\label{sec:methods}

To estimate the scientific impact of a noiseless ERD, we calculate the number of exoplanet detections and spectral characterizations during a two-year (exposure-time) survey optimized for exo-Earth candidate yield based on the LUVOIR-B mission concept \cite{luvoirfinalreport}. We adopt a multi-step process. First, we use the Altruistic Yield Optimization (AYO) tool to generate a list of observations optimized for exo-Earth candidates. Next, we use the ExoVista tool to generate planetary systems with planets of all types around nearby stars. Finally, we apply the exo-Earth-optimized observations from AYO to the ExoVista-generated planetary systems and simulate coronagraphic observations to calculate numbers of detected and spectrally characterized planets across the categories defined by \cite{Kopparapu18}.

\subsection{Exo-Earth-Optimized Observations}
\label{sec:eeoo}

We adopt two instrument concept scenarios for comparison. The first scenario adopts the LUVOIR-B mission concept parameters as presented by \cite{luvoirfinalreport}, using a deformable mirror-assisted vortex coronagraph (DMVC) with EMCCDs and an IFS for spectral characterization. The second scenario is identical to the first, but adopts an ERD in place of the EMCCD and IFS. For our baseline ERD, we assume the potential performance parameters of an ERD. Table \ref{tab:instruments} lists the performance parameters adopted for each scenario. We make relatively optimistic assumptions about the ERD, e.g., 99\% QE and 100\% fill factor, which have yet to be achieved by either TESs or MKIDs. As such, our estimates should serve as a reasonable upper limit on the potential scientific impact of an ERD.

\begin{table}[htb]
    \hspace{-0.5in}
    \begin{tabular}{l | l | l | l}
    \hline
    Parameter & EMCCD Model & ERD Model & Description \\
    \hline
    \hline
      &  &  & {\bf General Parameters} \\
    $\Sigma_T$               & 2 yrs & 2 yrs & Total exoplanet science time of the mission \\
    $\tau_{\rm slew}$        & 1 hr  & 1 hr  & Static overhead for slew and settling time \\
    $\tau_{\rm WFC}$         & $5\!\left(\frac{A_0\Upsilon_0}{A\Upsilon}\right)$ hrs & $5\!\left(\frac{A_0\Upsilon_0}{A\Upsilon}\right)$ hrs & Static overhead to dig dark hole \\
    $\tau'_{\rm WFC}$        & 1.1   & 1.1   & Multiplicative overhead to touch up dark hole \\
    $X$                      & 0.7   & 0.7   & Photometric aperture radius in $\lambda/D_{\rm LS}$$^a$ \\
    $\Omega$                 & $\pi\!\left(\frac{X\lambda}{D_{\rm LS}}\right)^2$ sr & $\pi\!\left(\frac{X\lambda}{D_{\rm LS}}\right)^2$ sr & Solid angle subtended by photometric aperture$^a$ \\
    $\zeta_{\rm floor}$      & 10$^{-10}$ & 10$^{-10}$ & Raw contrast floor \\
    $\Delta {\rm mag_{floor}}$ & 26.5  & 26.5 & Noise floor (faintest detectable point source at S/N$_d$) \\
    $T_{\rm contam}$         & 0.95  & 0.95  & Effective throughput due to contamination \\
    \hline
      &  &  & {\bf Detection Parameters} \\
    $\lambda_{\rm d,1}$      & 0.45 ${\rm \mu m}$ & 0.45 ${\rm \mu m}$ & Central wavelength for detection in SW coronagraph \\
    $\lambda_{\rm d,2}$      & 0.55 ${\rm \mu m}$ & 0.55 ${\rm \mu m}$ & Central wavelength for detection in LW coronagraph \\
    S/N$_d$                  & 7     & 7     & S/N required for detection (summed over both \\ & & & \hspace{0.25in} coronagraphs) \\
    $T_{\rm optical,1}$      & 0.17 & 0.17 & End-to-end reflectivity/transmissivity at $\lambda_{\rm d,1}$ \\
    $T_{\rm optical,2}$      & 0.39 & 0.39 & End-to-end reflectivity/transmissivity at $\lambda_{\rm d,2}$ \\
    $\tau_{\rm d,limit}$     & 2 months & 2 months & Detection time limit including overheads \\
    \hline
      &  &  & {\bf Characterization Parameters} \\
    $\lambda_c$              & 1.0 ${\rm \mu m}$ & 1.0 ${\rm \mu m}$ & Wavelength for characterization in LW coronagraph IFS \\
    S/N$_c$                  & 5     & 5     & Signal to noise per spectral bin evaluated in continuum \\
    R                        & 140    & 140    & Spectral resolving power \\
    $T_{\rm optical,c}$    & 0.23 & 0.39 & End-to-end reflectivity/transmissivity at $\lambda_c$ \\
    $\tau_c$                 & 2 months & 2 months & Characterization time limit including overheads \\
    \hline
      &  &  & {\bf Detector Parameters} \\
    $n_{\rm pix,d}$          & 8     & 8 & Number of pixels in photometric aperture at each $\lambda_d$ \\
    $n_{\rm pix,c}$          & 192   & 8 & Number of pixels per spectral bin in LW coronagraph \\ & & & \hspace{0.25in} IFS at $\lambda_c$ \\
    $\xi$                    & $3\!\times\!10^{-5}$ $e^-$pix$^{-1}$s$^{-1}$ & 0 & Dark current \\
    RN                       & 0 $e^-$pix$^{-1}$read$^{-1}$ & 0 & Read noise \\
    $\tau_{\rm read}$        & N/A   & N/A & Time between integrated reads \\
    CIC                      & $1.3\!\times\!10^{-3}$ $e^-$pix$^{-1}$frame$^{-1}$ & 0 & Clock induced charge \\
    $T_{\rm QE}$             & 0.9$^b$   & 0.99 & Raw QE of the detector at all wavelengths \\
    $T_{\rm read}$           & 0.75  & 1.0 & Effective throughput due to charge transfer, \\ & & & \hspace{0.25in} inefficiency, and cosmic ray mitigation \\
    \hline
    \end{tabular}
    \caption{Coronagraph-based mission parameters for our two detector scenarios. Parameters for the EMCCD model are based on \cite{Stark19}. \\ $^a\,D_{\rm LS}$ = diameter of Lyot stop projected onto primary mirror. \\ $^b$ Assumed for future detector performance by the LUVOIR mission concept \cite{luvoirfinalreport}.}
    \label{tab:instruments}
\end{table}

For each scenario, we adopted the same occurrence rates, exo-Earth candidate definition, albedo, and phase function as were used in the LUVOIR final report. We used the AYO code to run exo-Earth candidate yield calculations \cite{Stark14, Stark15, Stark19} and adopted observational requirements nearly identical to those used in the LUVOIR final report. Specifically, we required a minimum of six observations of each star to account for orbit determination, as well as an S/N=5, R=140 spectrum at 1.0 ${\rm\mu m}$ in the characterization phase to search for water on each exo-Earth candidate. We increased the spectral resolving power from the R=70 adopted in the LUVOIR final report to R=140, motivated in part by \cite{Damiano22} and \cite{Stark23}.

These yield calculations produced a list of target stars and exposure times for each of the instrument scenarios adopted, which formed the base of subsequent observation simulations. Because of the higher efficiency of our ERD, exposure times were shorter, and a greater number of planetary systems were generated for the ERD scenario (222 target stars for ERD compared with 168 for EMCCD+IFS). The EMCCD target list was nearly a subset of the ERD target list, but it included two stars that were not on the ERD list. (This is an artifact of AYO's apportionment of observation time to targets.)

The ExoVista code used to generate randomized planetary systems is limited in the host stars it can simulate because it does not currently support stellar spectrum models for $\log g > 5.25$, nor for $T_{\rm eff} < 3500$ K \cite{Kurucz}, so these stars are automatically excluded. These restrictions required the removal of 4 stars from the AYO observation plan for the EMCCD target sample and 7 stars from the ERD target sample. We do not expect this exclusion to significantly impact our final results, nor will it significantly change the result of the calibration test discussed in Section \ref{sec:calibrate}.

The LUVOIR survey design was divided into two phases \cite{luvoirfinalreport}. The first, herein called the ``detection phase,'' was intended to detect and identify exo-Earth candidates as well as their orbits. In this phase, each star is visited six times with epochs and exposure times calculated to maximize exo-Earth detections. To constrain orbital properties, a planet would have to be detected roughly three times at different phases; budgeting for $\sim$50\% detection efficiency, six detections were mandated for each star to approximately account for this requirement \cite{luvoirfinalreport}.

The second phase, herein called the ``characterization phase,'' was intended to revisit the most promising exo-Earth candidates (EECs) for spectroscopic follow-up. In this phase, a subset of the target list with promising EECs (38 stars for ERD and 28 stars for EMCCD+IFS) would be visited once each at an appropriate epoch to observe the EEC at an optimized orbital phase. The exposure times in this phase were calculated to achieve the desired $S/N=5$ for a robust detection of water at 1.0 ${\rm\mu m}$ \cite{Stark23}.

\subsection{ExoVista Planetary Systems}
\label{sec:systems}

For each star in the target list generated above, we produced synthetic planetary systems using the open-source ExoVista tool \cite{exovista}\footnote{https://github.com/alexrhowe/ExoVista}. In this paper, we used a build of ExoVista Version 2.1 customized to replicate the AYO outputs. ExoVista 2 is a port of the original exoVista tool from IDL/C to Python/C++. Additional features are in development and will be discussed in a forthcoming paper.

A custom build of ExoVista was used because the AYO yield tool does not contain the same fidelity as ExoVista. As a result, the exposure times generated by AYO are not optimized for the specific planetary systems generated by ExoVista. Ideally, we could feed the ExoVista systems into AYO to properly optimize observations. However, this is currently out of the scope of our investigation. To rectify this issue, we temporarily simplified the ExoVista models to better approximate the assumptions made by AYO. This procedure manifested in two major changes.

First, AYO assumes a flat geometric albedo of 0.2 for planets $<1.4\,R_\oplus$ and 0.5 for planets $>1.4\,R_\oplus$. The standard ExoVista model uses geometric albedo spectra for various planet types. Because of the assumptions used in assigning planet types, many regions of the parameter space as computed by ExoVista contain planets that are brighter or (more often) fainter than the AYO model, with commensurate effects on their detectability. In particular, most of the cold super-Earth planets are assumed to have the spectrum of Mars, which has an albedo of 0.1 in the V-band, making them significantly less detectable. For this paper, we changed the geometric albedos assigned by ExoVista to match those of the AYO model.

Second, ExoVista draws individual planets randomly from the entire occurence rate distribution in radius-flux space. While this is the most intuitive method, AYO calculates the expectation value of the number of planets per bin in the parameter space for each planet type, and a fully random distribution will not recreate the expectation values exactly. Therefore, we adjusted ExoVista's random planet draw to randomly add and remove planets of each type to match the AYO calculation. In other respects such as the input occurrence rate table and the stability criterion used, the planetary population we generated with ExoVista already matched those of the AYO model.

These changes leave a degree of randomness in the ExoVista-generated systems, as the randomly-generated planets will not match the expectation values computed by AYO exactly on a star-by-star basis. These differences should be minimized by summing over the entire target list, but small differences will persist. We discuss these variations and the methods we used to mitigate them in Section \ref{sec:calibrate}.

Finally, we note that AYO adopts a circularly symmetric analytic exozodi model, whereas ExoVista adopts a physically-motivated model using Mie theory and realistic forward scattering (see \cite{Stark14}, Eq. C4). To obtain reasonable agreement between the two codes for this paper, we normalized ExoVista's disk model to match AYO's disk brightness measured at the Earth equivalent insolation distance (EEID) along the disk major axis, but retained the Henyey-Greenstein (HG) scattering phase functions \cite{HG41} used by ExoVista. This change provided a balance between more realistic physics and the need to calibrate the code to AYO. With this normalization, the realistic scattering properties will not significantly change the detectability of planets near quadrature, but it is likely to increase the detectability of planets in gibbous phase. With the forward-scattering implemented in ExoVista, the far side of the disk is fainter (due to weaker back-scattering) than the near side. Since planets in gibbous phase appear in the far side of the disk, less exozodiacal noise will be mixed with their signal, making them more detectable.

\subsection{Observation simulations}
\label{sec:obs}

Simulating the observations of the planetary systems with the LUVOIR-B telescope and instrument model generally follows the procedure laid out in \cite{Stark19}, adapted for the output format reported by ExoVista. Constants in these formulae specific to each observing mode are listed in Table \ref{tab:instruments}. The signal photon count rate from the planet received by the telescope per spectral resolving element is given by
\begin{equation}
    {\rm CR_p} =  F_0\times 10^{-0.4{\rm m}_V} \frac{f_p}{f_*} \Upsilon_c AT \Delta\lambda,
\end{equation}
where $F_0$ is the zero point of the magnitude system, m$_V$ is the apparent V-band magnitude of the star, $\frac{f_p}{f_*}$ is the contrast ratio between the planet and the star, $A$ is the collecting area of the telescope, $\Upsilon_c$ is the core throughput of the coronagraph integrated over the photometric aperture (a function of the planet-star separation), $T$ is the end-to-end instrument throughput (other than the coronagraph's core throughput), and $\Delta\lambda$ is the bandpass of the resolving element. We note that $T$ is separated into multiple throughput components in Table \ref{tab:instruments}, such as QE.

The background noise for exoplanet observations includes coronagraphic leakage from the star, local zodiacal light, exozodiacal light, and, in the case of the EMCCD, instrument noise (dark current and clock-induced charge). In reality, there are also galactic and extragalactic astrophysical backgrounds. However, we assume that these background sources can be minimized by appropriate target selection or observation timing, which is beyond the scope of this paper.

The background for stellar leakage is given by
\begin{equation}
    {\rm CR_{b,*}} = F_0\times 10^{-0.4{\rm m}_V} \zeta\cdot {\rm PSF_\Omega} AT \Delta\lambda.
\end{equation}
Here, $\zeta$ is the coronagraphic contrast ratio, and ${\rm PSF_\Omega}$ is the fraction of the star's point-spread function (PSF) in the photometric aperture. This equation is equivalent to Eq. 10 in \cite{Stark14}, except replacing the approximation of ${\rm PSF_\Omega} \approx {\rm PSF_{peak}}\Omega$ with the direct numerical integral of the PSF over the photometric aperture. Both of these parameters are functions of planet-star separation and taken together are analogous to the $\Upsilon_c$ term for the planet.

The background for zodiacal light is given by
\begin{equation}
    {\rm CR_{b,zodi}} = F_0 \times 10^{-0.4z} \Omega T_{\rm sky} AT \Delta\lambda,
\end{equation}
where $z$ is the magnitude surface brightness of the zodical light (assumed to be a constant for the purpose of this paper), $\Omega$ is the solid angle of the photometric aperture (thus determining the total contamination reaching the detector), and $T_{\rm sky}$ is the effective throughput of the coronagraph when convolved with a diffuse background (a function of planet-star separation, but different from $\Upsilon_c$). The exozodiacal count rate (assuming an optically-thin disk) obeys the same formula, except that the exozodiacal surface brightness, $z$, becomes a function of planet-star separation and the density of the disk provided by the ExoVista simulation.

Finally, the instrumental background applies only to the EMCCD scenario, since the noise levels for our ERD are considered to be negligible. It is given by
\begin{equation}
    {\rm CR_{b,i}} = n_{\rm pix}(\xi + {\rm RN}^2/\tau_{\rm read} + 6.73\,{\rm CIC}\cdot{\rm CR_{sat}}).
\end{equation}
Here, $n_{\rm pix}$ is the number of pixels in the PSF (always the same except when the IFS is used), $\xi$ is the dark current, RN is the read noise, which we assume to be negligible, CIC is the clock-induced charge, and CR$_{\rm sat}$ is the read rate of the EMCCD sensor. We set ${\rm CR_{sat}} = 10{\rm CR_p}/n_{\rm pix}$ s$^{-1}$, such that the sensor is unlikely to detect more than one photon per count reported even for sources ten times brighter than an exoEarth at quadrature. The coefficient of 6.73 in Equation (4) results from assuming a Geiger efficiency of 99\%, ensuring that less than 1\% of photons are lost due to operating the EMCCD in photon counting mode \citep{Stark19}.

With these backgrounds calculated, we can then compute the signal-to-noise ratio of an AYO-optimized observation with integration time $\tau$ with
\begin{equation}
    ({\rm S/N})^2 = \frac{1}{\tau}\left(\frac{\rm CR_p^2}{\rm CR_p + 2CR_{b,tot}}\right).
\end{equation}

The four parameters that are functions of star-planet separation ($\Upsilon_c$, $\zeta$, ${\rm PSF}_\Omega$, and $T_{\rm sky}$) were computed numerically from the azimuthally-averaged DMVC coronagraph performance model presented by \cite{Stark19}. We did not apply a sharp cutoff at the inner working angle (IWA), instead applying the coronagraph throughput and stellar leakage terms, which result in S/N falling rapidly within the IWA. We set the outer working angle (OWA) to $60\lambda/D$ to match AYO's assumptions.


\subsubsection{Calibrating ExoVista to AYO}
\label{sec:calibrate}

Our analysis in this paper relies on applying AYO-optimized exposure times to ExoVista-generated planetary systems with these simulated observations. Therefore, it is critical that we verify the exposure times are consistent. To do this, we chose three fiducial stars from the target list of different spectral types, then simulated systems with an Earth twin around them, placed at quadrature.\footnote{We modeled the Earth twins with zero eccentricity so that the eccentric anomaly would equal the mean anomaly, and thus the longitude for quadrature would be exact.} We then compared the photon count rates for the signal and each of the noise terms between ExoVista and the AYO model, using the AYO exposure times. (Since our exozodiacal model was normalized to this case, it would also be consistent with the AYO model.)

If the photon counts and signal-to-noise ratios from ExoVista and AYO matched for an Earth twin at quadrature (this being the model for which the exposure times were computed), it would demonstrate that ExoVista had accurately recreated the AYO planetary system and flux models. We found good agreement between the two for our test cases.

The next step was to verify the number of planets detected and number of spectra taken in the target list between ExoVista and AYO. This step was less exact because of irreducible differences between the design of the two codes. While AYO focuses on the statistical expectation value of yields, ExoVista randomly distributes discrete planets. This random generation of planets by ExoVista means that each planet type may have different detectability rates between simulations. Further, even in the event of identical planets, the random distribution of orbital elements and epochs and the random assignment of planets of the same instellation to stars of different spectral types means that the planets may be observable or not in the coronagraph field in different simulations, especially those that orbit at distances where they may be excluded due to the outer working angle. Additionally, the orbital epochs at which the planets are observed will be {\it a priori} random, and some planets with a high detection probability may not be detected because of an unfavorable orbital configuration.


To address these sources of uncertainty, we modeled the list of observations provided by the AYO tool for each scenario multiple times with the starting epoch offset by 10 days each time. The median value of the number of planets detected of each type (after accounting for the stars that were excluded from the ExoVista target list, as discussed in Section \ref{sec:eeoo}) is the value we adopted as the number of predicted detections.

The AYO tool predicts the number of planets detected by a LUVOIR-style survey in the EMCCD scenario specifically. (Note that this includes only the number of distinct detections, not characterization spectra.) Thus, we can calibrate our ExoVista-based method by comparing the results of our simulation in that scenario to the AYO predictions. However, for this purpose, our simulated survey was not large enough to obtain clear statistics, so we repeated the survey 100 times to improve these statistics, randomly generating a new set of planets and disks for the stellar targets each time and averaging over the 100 simulations. To speed these calculations, we degraded the precision of the orbital integration, including deactivating transit detections, since precise ephemerides were not needed, only a large distribution. We also degraded the precision of the disk flux computation; this should only have a small effect on the results because high disk densities that require higher-precision calculations mainly occur in the centers of disks inside the inner working angle, and in the relatively small population of edge-on disks.

After exclusions by the code, this larger ensemble of simulated surveys included 15890 planetary systems across the 100 simulations. We then averaged the exoplanet detections in the ``detection phase'' of the surveys to compare with the AYO yields. We discuss our methods for counting of the number of detected planets in greater detail in Section \ref{sec:counting}. The detection counts predicted with the AYO tool and for ExoVista are shown in Table \ref{tab:ayocal}.

In principle, the averaged simulated surveys using ExoVista scenes should return the same number of planets of each type detected as the AYO calculations. In practice, the numbers will not be equal for several reasons. These will include the inherent randomness in our generated population of systems, but the most likely factor is ExoVista's non-axisymmetric disk model, which renders planets significantly more detectable in gibbous phase while embedded in the fainter far side of the disk. Because of this, we predict that any discrepancies would lean toward ExoVista counting more planets than AYO. Additionally, ExoVista may treat edge-on disks more faithfully than AYO, which assumes a simple $1/r^2$ brightness profile compared with ExoVista's Mie theory model.

In our results, we find that 10 of the 15 planet types we counted (as described in Section \ref{sec:counting}) had ExoVista yields within 8\% of AYO's, and the remaining five types (hot sub-Neptunes, Neptunes, and Jupiters, and warm and cold sub-Earths) were all overcounted, consistent with expectations. Further, all five of the overcounted planet types fall into two categories that are the most marginal for detection and thus the most model-sensitive in their resulting yields. These are hot planets, which are more likely to be lost in the glare of the inner parts of the disk; and sub-Earths, which are smaller than the target size of planets to be detected by the survey. Therefore, we consider our exoVista approach to calculating yields adequately benchmarked to the AYO approach given the codes' fundamental differences.



\begin{table}[htb]
    \centering
    \begin{tabular}{l | c | c | c | c | c}
    \hline
    Planet Type & Sub-Earth & Super-Earth & Sub-Neptune & Neptune & Jupiter \\
    AYO / ExoVista & & & & & \\
    \hline
    Hot  & 43.2 / 41.9 & 50.9 / 51.3 &  51.4 / 64.0  &  8.6 / 10.4 &  7.9 / 8.9  \\
    Warm & 17.1 / 20.7 & 33.9 / 33.8 &  36.3 / 36.6  & 12.1 / 12.1 & 14.0 / 12.9 \\
    Cold &  4.5 / 6.5  & 42.2 / 42.6 & 102.5 / 97.8  & 71.9 / 67.8 & 72.0 / 74.7 \\
    \hline
    \end{tabular}
    \caption{Comparison of the average detection yield across 100 simulated HWO-style surveys of planetary systems generated by ExoVista compared with the expected values predicted by AYO. The left number of each pair indicates the yield from AYO for a specific planet type, with the truncated target list of 164 stars. The right number indicates the same result from ExoVista. The two approaches broadly agree.}
    \label{tab:ayocal}
\end{table}

\subsubsection{Counting Detections and Spectra}
\label{sec:counting}

To determine the expected survey yields, we simulated both phases of the survey (detection and characterization) using the optimized observations from AYO. The AYO tool produces a series of observations of target stars with epochs and exposure times for the detection phase, as well as additional observations for the characterization phase. Much like our verification of exo-Earth candidate observations, we used the scenes produced by ExoVista combined with the coronagraph model described in Section \ref{sec:obs} to emulate what a LUVOIR-style observatory would see in each observation. For each planet in the field of view, we considered observations with ${\rm S/N}>7$ integrated over the bandpass to be valid planet detections, subject to the OWA. For spectral characterizations, our target was ${\rm S/N} = 5$ at R=140, averaged over the bandpass (if the instrument was collecting spectral information).

Note that we make no changes to the CONOPS of the LUVOIR-B survey model that we baselined for a prospective HWO survey. Exo-Earth candidates, which are used to select targets for spectral characterization, still require six detections to determine their orbits and subsequent observation times. Many of the other planets in these systems will have fewer than six detections, leaving their orbits less constrained or unconstrained. These are the planets in which we are interested in this study, but we do not require them to have established orbits, because we are not specifically targeting them for spectral characterization, but only counting the number of incidental spectra that we get ``for free'' over the course of the survey.

For the EMCCD scenario, detections are counted during both phases of the survey. However, spectra are counted only during the characterization phase, when the IFS is in use. In contrast, our ERD scenario resolves spectral information at all times. There is no separate imaging mode, so both detections and spectra are counted during both phases of observations for the ERD.

In each case, we tracked the number of detections and spectra by planet type, binned according to the classification of \cite{Kopparapu18}. This classification includes 15 planet types based on five bins in radius and three bins in instellation. In radius, the planets are divided into sub-Earths ($R<1\,R_\oplus$), super-Earths ($1\,R_\oplus<R<1.75\,R_\oplus$), sub-Neptunes ($1.75\,R_\oplus<R<3.5\,R_\oplus$), Neptunes ($3.5\,R_\oplus<R<6\,R_\oplus$), and Jupiters ($R>6\,R_\oplus$). In instellation, the planets are divided into hot, warm, and cold types, roughly corresponding to their being interior to the habitable zone, in the HZ, and exterior to the HZ, respectively. However, the \cite{Kopparapu18} classification did not extend the hot and cold types to the full range of planets in the ExoVista model, so planets at the extremes of semi-major axes (roughly interior to $\sim$0.07 AU and exterior to $\sim$18 AU, scaled with stellar luminosity) were not included in these counts. We adopted the exo-Earth candidate definition from the LUVOIR and HabEx final reports \cite{luvoirfinalreport,habexfinalreport}. EECs are a subset of other planet types, overlapping with both the warm sub-Earth and warm super-Earth categories.

We applied the AYO observations to the ExoVista planetary systems, counting planet detections and spectra that met the respective signal-to-noise thresholds. Counting the number of planet detections in each scenario was a simple matter of counting each planet in the simulation that appeared in at least one observation with high enough S/N, combining the six visits per star in the detection phase with the additional one visit per star in the characterization phase (in the smaller characterization sample). Likewise, for spectral characterization, we counted each planet that had a successfully extracted spectrum in at least one observation---in the characterization phase only for the EMCCD scenario, but in both phases for the ERD scenario.

As discussed in Section \ref{sec:calibrate}, we drew many possible sets of observations for each scenario, including multiple possible characterization samples for each set of detection observations. For the detection phase, we drew the set of observations 73 times, offset by 10 days each time. In other words, we effectively simulated 73 detection phases with their epochs randomized across a span of two years, counting the number of detections and characterizations in each simulation. We then adopted the median number of planets detected in each bin (summed across the full target list) as our final number of detections.

An additional source of randomness occurs in the choice of the sample used during the characterization phase of the observations, which is a subsample of approximately one sixth of all of the observed systems. This sample size is based on the expected yield of a HWO-style mission, and in the final survey, these targets would be selected based on which ones were found to have the most promising exo-Earth candidates for characterization. In this study, we draw them randomly. To improve the statistics, we randomly drew the smaller characterization sample nine times for each offset in observing times, again taking the median result as our number of predicted spectra. This mitigates the problem that we do not know \textit{a priori} which systems will be selected for spectral characterization. Within each planet type, we again adopted the median values of detections and characterizations for our adopted yields for the LUVOIR-B survey design as a whole.

We note that in the ERD scenario, the vast majority of spectra are taken during the \emph{detection phase}, which has many more target stars and more visits per target than the characterization phase. The primary goal of the characterization phase is to revisit the detected exo-Earth candidates with optimal observing conditions. New incidental spectra taken in this phase are added to the total number, although any new unique detections are limited to those planets that by chance were not successfully observed during the detection phase.

\section{Results and Discussion}
\label{sec:results}

\begin{figure}[!ht]
    \hspace{0.55in}
    \includegraphics[width=0.7\textwidth]{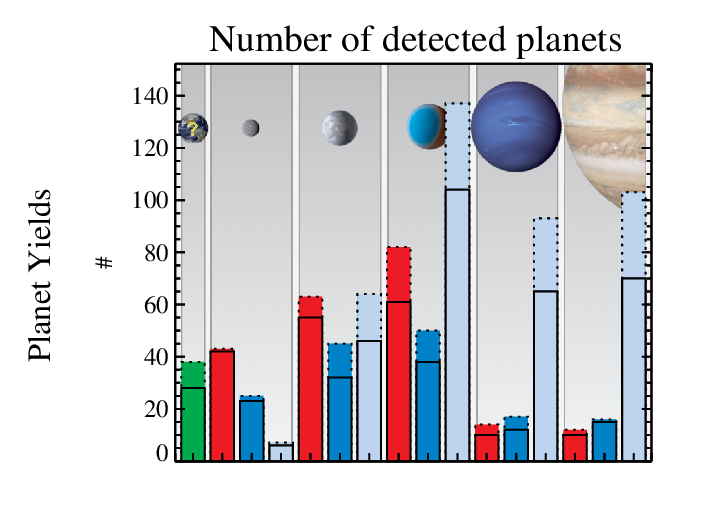}
    \caption{The number of planets detected in our simulated survey in the EMCCD scenario (solid) and the ERD scenario (dotted) for each planet type. In both cases, the numbers reflect our results for a single survey with a random realization of orbital positions, as described in Section \ref{sec:counting}, rather than an ensemble of surveys. The difference between the scenarios is mostly due to the higher throughput and lower noise of our baseline ERD (based on a TES array), which allow for a larger target list. The numbers of exo-Earth candidates (shown in green) are taken from the survey design numbers for the size of the characterization target list. (EECs overlap with both the warm sub-Earth and warm super-Earth counts.)}
    \label{fig:detections}
\end{figure}

The numbers of unique detected planets for the various planet types in both of our scenarios are plotted in Figure \ref{fig:detections}. The exo-Earth candidates are marked in green, with the planet types from \cite{Kopparapu18} marked by the other bars. (However, the yields of EECs are taken directly from the survey design instead of the counts---specifically, from the size of the characterization target list.) The solid outlines mark the yields from the EMCCD scenario, and the dotted outlines mark the yields from the ERD scenario. For this figure, we did not use an ensemble of surveys with offsets in starting epochs as in Section \ref{sec:calibrate}, but we instead used a single survey with a random realization of orbital positions for each scenario, computed at a higher fidelity.

For the detection phase, the difference between the two scenarios (and the two observational instruments) is mainly determined by the size of the target list. A survey such as that planned for HWO will necessarily be optimized to detect and characterize exo-Earth candidates, with any other detections being incidental. Thus, the greater throughput and lower noise levels of the ERD are leveraged for shorter exposure times and a larger target list. This is why the number of exo-Earth candidates detected (green), the metric the survey was designed around, is greater in the ERD scenario (38 for ERD versus 28 for EMCCD).

With observations optimized for detecting EECs at the same S/N in both scenarios, the ERD observing program should not be significantly better or worse than the EMCCD program for detecting other planet types. We would expect the difference in detected planet counts to arise from the size of the target list and to be proportionately greater for all planet types in the ERD scenario. This appears to be the case for most planet types with enough detections to provide good statistics. The exceptions are the sub-Earth planets, which have near-identical numbers of detected planets in the two scenarios. We attribute this to the random variation in planet-by-planet detectability for the planets generated by ExoVista.

\begin{figure}[!ht]
    \includegraphics[width=0.99\textwidth]{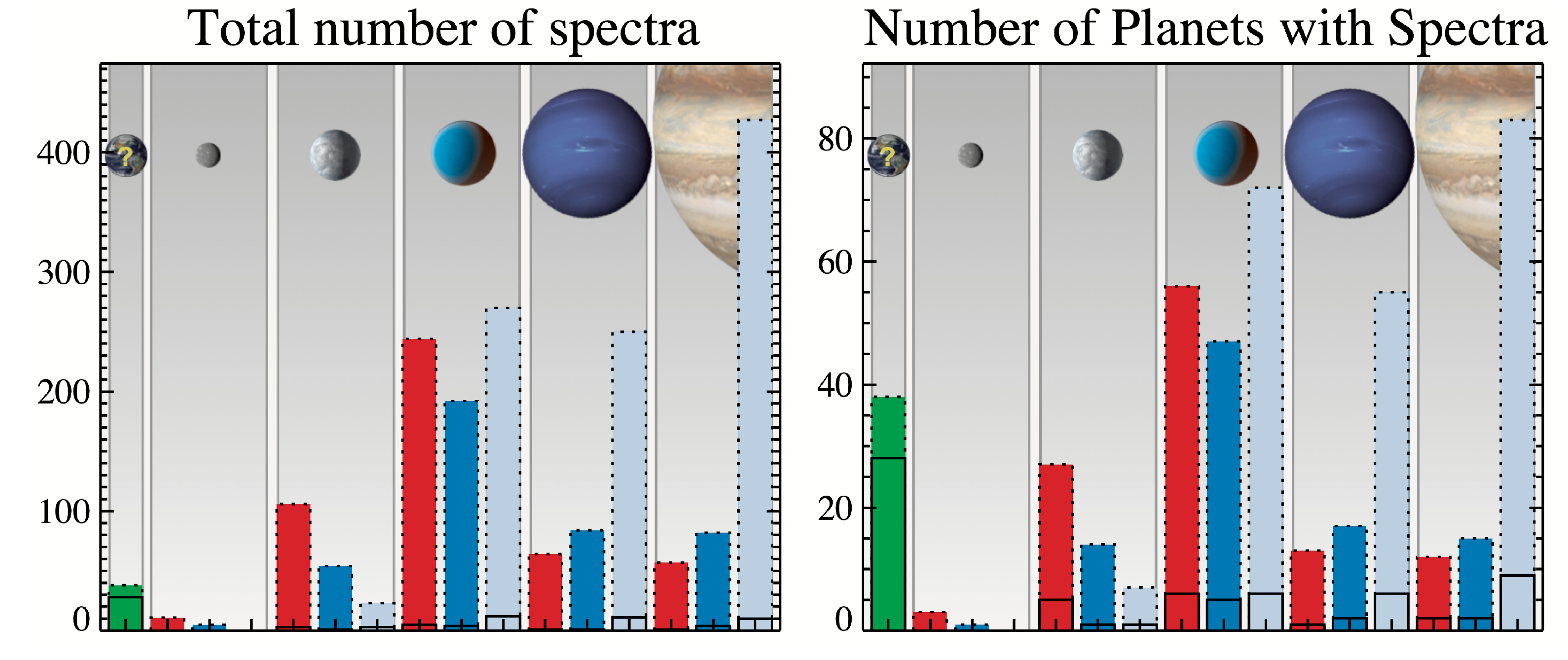}
    \caption{The number of spectra (left) and number of unique planets with spectra (right) taken by our simulated survey in the EMCCD scenario (solid) and ERD scenario (dotted) for each planet type. The designed yield of the survey of exo-Earth candidate spectra is shown in green and is equal to the size of the characterization target list in each case. The ERD scenario yields many more incidental spectra because of its ability to take spectra at all times, including during the detection phase.}
    \label{fig:counts}
\end{figure}

The total number of spectra taken and the number of unique planets with spectra in the two scenarios are plotted by planet type in Figure \ref{fig:counts}. Again, the solid outlines mark the yields from the EMCCD scenario, and the dotted outlines mark the yields from the ERD scenario. The ERD scenario obtains far more spectra than the EMCCD scenario.

The LUVOIR-B detection strategy was designed to detect planets and measure their orbits. In the case of an EMCCD, this provides significant incidental data about the architectures of these systems (hence the large number of incidental detections of all types of planets). However, this detection phase provides relatively \emph{little} information about the compositions of the planets. The follow-up characterization phase, in which systems with promising exo-Earth candidates are selected for spectroscopic characterization, does result in the collection of incidental spectra of other types of planets, but the sample is limited by the relatively small number of exoEarth candidates. Further, this sample will potentially be biased, given that planet types such as warm Jupiters that would preclude the existence of EECs will not be sampled.

Other planets in the same systems as EECs will often be fainter if farther from the star, or suffer from higher backgrounds if closer to the star, and likewise will often suffer lower signal-to-noise ratios than the primary targets. During spectral characterization, these planets will also typically have less favorable orbital alignments, since observations will not be planned to place them near quadrature. Thus, it is not surprising that a survey designed to characterize 28 exo-Earth candidates (in the EMCCD scenario) will yield just 48 incidental spectra of other planets (compared with the relative abundance of detections in the same scenario).

The advantage of a noiseless ERD over an EMCCD+IFS is partly in its higher throughput and lower noise, which allow for shorter exposure times, resulting in a significant $\sim$30\% increase in exo-Earth candidate yield. However, an even greater advantage is in its ability to take spectra at all times. While an EMCCD+IFS could do this as well, the detector’s noise properties would reduce the exo-Earth yield of the mission. This capability results in $\sim$6 times the yield of incidental spectra as a fraction of the ERD target list (making it $\sim$8 times the yield when compared with the smaller EMCCD target list). This much larger sample will be drawn more uniformly from the field population of exoplanets as a whole (within the bounds of the coronagraph window), as opposed to only EECs, providing a fuller picture of the atmospheric compositions of the exoplanet population. ERDs also have another benefit, in that they can more easily filter out cosmic ray contaminants from the data based on timing coincidences.


Our predicted planet detection yields for both instrument scenarios may be optimistic in several ways. First, some astrophysical assumptions may be optimistic. In particular, the flat geometric albedos assumed by AYO for terrestrial planets may be reasonable for water worlds, but are likely to be too high for rocky planets. Additionally, ExoVista assumes a Lambertian phase function for non-phase-resolved albedos. The actual phase functions of Solar system planets are significantly fainter than a Lambertian near quadrature (see e.g. \cite{Sudarsky05}), which will further reduce detectability. Future work is needed to investigate these uncertainties and further refine the yield estimates presented in this paper.

We also made several optimistic assumptions about instrument performance. For both the ERD array and EMCCD+IFS scenarios, we adopted a high QE at all wavelengths. For the EMCCD+IFS, we neglected the degradation of the detector over time due to cosmic rays\cite{CCDrad} as well as the use of the high-gain amplifier. For the ERD, we assumed a large fill-factor and zero noise. While all of these assumptions may prove achievable, they have yet to be demonstrated. In addition, the mK operating temperature of an ERD may require thermal isolation that mandates additional optics, reducing its effective throughput. We also note that alternative detectors to EMCCDs, such as a Skipper CCD, could be paired with an IFS, which would improve the science yield of the IFS scenario. As such, we regard our estimates of the ERD science impact as upper limits. Our work highlights the importance of future detailed trade studies of all possible detector technologies for HWO.

Finally, there may be unexplored flexibility in the overall survey design we studied. Our analysis adopted the LUVOIR baseline observation strategy, in which multi-epoch observations are first conducted to measure orbits and select the EECs that are most promising for spectral characterization follow-up. This operations concept was developed in part due to the separate imaging and spectral characterization modes. With an ERD, no such separate modes exist and there is no penalty for taking spectra. As such, it is no longer clear whether the LUVOIR baseline observing strategy is ideal, or whether deeper observations designed to obtain spectra from the start would be fruitful. Future work should re-examine the observing strategy of an ERD-based corongraphic instrument.

\section{Conclusion}
\label{sec:conclusion}

We have investigated the effect of a noiseless energy-resolving detector on a survey for potentially-habitable planets such as that planned for the future Habitable Worlds Observatory (HWO). Baselining the possible performance parameters of a TES array, we quantified the advantage that an ERD could potentially provide over the EMCCD+IFS baselined by the LUVOIR and HabEx studies. The higher throughput and lower noise of our baseline ERD allows for shorter exposure times and thus for observing a larger target list, giving it the potential to increase the detection yields of exo-Earth candidates by up to $\sim$30\%. The potential for incidental exoplanet science on other detected planets is even greater because ERDs take spectra at all times. This gives them the potential to observe spectra for hundreds of additional exoplanets ``for free'' ($\sim$8 times as many as the targeted characterization sample in the EMCCD scenario).

We verified the ExoVista tool to accurately generate randomized exoplanet systems and simulated observations that are consistent with the earlier predictions of the Altruistic Yield Optimization (AYO) tool for a LUVOIR-style survey. ExoVista provides a more physically-detailed simulation of planetary systems and observations, which can be used for higher fidelity survey modeling.

\section*{Disclosures}

The authors have no relevant conflicts of interest to disclose.

\section*{Code, Data, and Materials Availability}

The ExoVista code that used in this paper is available at https://github.com/alexrhowe/ExoVista.

The target lists and AYO-generated observations used in this paper are available at \\ https://www.starkspace.com/permanent/Howe-TES\_yields.zip.

\section*{Acknowledgements}

The authors acknowledge productive conversations with Bernard Rauscher and Patrick Morrissey that helped improve this manuscript. ARH acknowledges support by NASA under award number 80GSFC21M0002 through the CRESST II cooperative agreement. This work was partially supported by the GSFC Exoplanet Spectroscopy Technologies (ExoSpec), which is part of the NASA Astrophysics Science Division’s Internal Scientist Funding Model. The authors thank NASA Goddard’s Internal Research and Development program.

\bibliography{refs}

\begin{thebibliography}{10}

\bibitem{ASTRO2020}
{National Academies of Sciences, Engineering, and Medicine}, {\em {Pathways to
  Discovery in Astronomy and Astrophysics for the 2020s}}  (2021).

\bibitem{ErtelExoZ}
S.~{Ertel}, D.~{Defr{\`e}re}, P.~{Hinz}, {\em et~al.}, ``{The HOSTS Survey for
  Exozodiacal Dust: Observational Results from the Complete Survey},'' {\em
  \aj} {\bf 159}, 177  (2020).

\bibitem{habexfinalreport}
B.~S. {Gaudi}, S.~{Seager}, B.~{Mennesson}, {\em et~al.}, ``{The Habitable
  Exoplanet Observatory (HabEx) Mission Concept Study Final Report},'' {\em
  arXiv e-prints} , arXiv:2001.06683  (2020).

\bibitem{luvoirfinalreport}
{The LUVOIR Team}, ``{The LUVOIR Mission Concept Study Final Report},'' {\em
  arXiv e-prints} , arXiv:1912.06219  (2019).

\bibitem{CCDrad}
P.~Morrissey, L.~K. Harding, N.~L. Bush, {\em et~al.}, ``{Flight photon
  counting electron multiplying charge coupled device development for the Roman
  Space Telescope coronagraph instrument},'' {\em Journal of Astronomical
  Telescopes, Instruments, and Systems} {\bf 9}(1), 016003  (2023).

\bibitem{Roman}
``{Nancy Grace Roman Space Telescope Simulations, Spacecraft and Instrument
  Parameters}.'' Webpage,
  \url{https://roman.ipac.caltech.edu/sims/Param_db.html}.
\newblock (Accessed: 16 April 2024).

\bibitem{IFSref}
T.~D. Groff, N.~Zimmerman, M.~J. Rizzo, {\em et~al.}, ``{Development of the
  WFIRST CGI integral field spectrograph},'' in {\em Techniques and
  Instrumentation for Detection of Exoplanets IX},  S.~B. Shaklan, Ed.,  {\bf
  11117}, 111170D, International Society for Optics and Photonics, SPIE
  (2019).

\bibitem{Rauscher16}
B.~J. {Rauscher}, E.~R. {Canavan}, S.~H. {Moseley}, {\em et~al.}, ``{Detectors
  and cooling technology for direct spectroscopic biosignature
  characterization},'' {\em Journal of Astronomical Telescopes, Instruments,
  and Systems} {\bf 2}, 041212  (2016).

\bibitem{MKIDref}
J.~{Gao}, {\em {The Physics of Superconducting Microwave Resonators}}.
\newblock PhD thesis, California Institute of Technology  (2008).

\bibitem{TESref}
K.~Irwin and G.~Hilton, {\em Transition-Edge Sensors}, 63--150.
\newblock Springer Berlin Heidelberg, Berlin, Heidelberg  (2005).

\bibitem{beaumont2023a}
S.~Beaumont, J.-M. Lauenstein, J.~S. Adams, {\em et~al.}, ``Effect of space
  radiation on transition-edge sensor detectors performance,'' {\em IEEE
  Transactions on Applied Superconductivity} {\bf 33}(5), 1--6  (2023).

\bibitem{beaumont2023b}
S.~Beaumont, J.~S. Adams, S.~R. Bandler, {\em et~al.}, ``Long term performance
  stability of transition-edge sensor detectors,'' {\em IEEE Transactions on
  Applied Superconductivity} {\bf 33}(5), 1--5  (2023).

\bibitem{deVisser21}
P.~J. {de Visser}, S.~A.~H. {de Rooij}, V.~{Murugesan}, {\em et~al.},
  ``{Phonon-Trapping-Enhanced Energy Resolution in Superconducting
  Single-Photon Detectors},'' {\em Physical Review Applied} {\bf 16}, 034051
  (2021).

\bibitem{Techport}
``{Ultra-high Efficiency Noiseless Quantum Sensors for HWO and QIS}.'' Webpage,
  \url{https://techport.nasa.gov/view/146757}.
\newblock (Accessed: 23 April 2024).

\bibitem{Kopparapu18}
R.~K. {Kopparapu}, E.~{H{\'e}brard}, R.~{Belikov}, {\em et~al.}, ``{Exoplanet
  Classification and Yield Estimates for Direct Imaging Missions},'' {\em \apj}
  {\bf 856}, 122  (2018).

\bibitem{Stark19}
C.~C. {Stark}, R.~{Belikov}, M.~R. {Bolcar}, {\em et~al.}, ``{ExoEarth yield
  landscape for future direct imaging space telescopes},'' {\em Journal of
  Astronomical Telescopes, Instruments, and Systems} {\bf 5}, 024009  (2019).

\bibitem{Stark14}
C.~C. {Stark}, A.~{Roberge}, A.~{Mandell}, {\em et~al.}, ``{Maximizing the
  ExoEarth Candidate Yield from a Future Direct Imaging Mission},'' {\em \apj}
  {\bf 795}, 122  (2014).

\bibitem{Stark15}
C.~C. {Stark}, A.~{Roberge}, A.~{Mandell}, {\em et~al.}, ``{Lower Limits on
  Aperture Size for an ExoEarth Detecting Coronagraphic Mission},'' {\em \apj}
  {\bf 808}, 149  (2015).

\bibitem{Damiano22}
M.~{Damiano} and R.~{Hu}, ``{Reflected Spectroscopy of Small Exoplanets II:
  Characterization of Terrestrial Exoplanets},'' {\em \aj} {\bf 163}, 299
  (2022).

\bibitem{Stark23}
C.~C. {Stark}, N.~{Latouf}, A.~M. {Mandell}, {\em et~al.}, ``{Optimized
  Bandpasses for the Habitable Worlds Observatory ExoEarth Survey},''
  (submitted).

\bibitem{Kurucz}
F.~{Castelli} and R.~L. {Kurucz}, ``{New Grids of ATLAS9 Model Atmospheres},''
  in {\em Modelling of Stellar Atmospheres},  N.~{Piskunov}, W.~W. {Weiss}, and
  D.~F. {Gray}, Eds.,  {\bf 210}, A20  (2003).

\bibitem{exovista}
C.~C. {Stark}, ``{ExoVista: A Suite of Planetary System Models for Exoplanet
  Studies},'' {\em \aj} {\bf 163}, 105  (2022).

\bibitem{HG41}
L.~G. {Henyey} and J.~L. {Greenstein}, ``{Diffuse radiation in the Galaxy.},''
  {\em \apj} {\bf 93}, 70--83  (1941).

\bibitem{Sudarsky05}
D.~{Sudarsky}, A.~{Burrows}, I.~{Hubeny}, {\em et~al.}, ``{Phase Functions and
  Light Curves of Wide-Separation Extrasolar Giant Planets},'' {\em \apj} {\bf
  627}, 520--533  (2005).

\end{thebibliography}
\bibliographystyle{spiejour}   

\listoffigures
\listoftables

\end{spacing}
\end{document}